# An Algorithmic Approach for Identifying Critical Distance Relays for Transient Stability Studies

Ramin Vakili, *student member, IEEE*, Mojdeh Khorsand, *member, IEEE*, and Vijay Vittal, *Fellow, IEEE,* Bill Robertson, Philip Augustin, *member, IEEE*

*Abstract*--After major disturbances, power system behavior is governed by the dynamic characteristics of its assets and protection schemes. Therefore, modeling protection devices is essential for performing accurate stability studies. Modeling all the protection devices in a bulk power system is an intractable task due to the limitations of current stability software, and the difficulty of maintaining and updating the data for thousands of protection devices. One of the critical protection schemes that is not adequately modeled in stability studies is distance relaying. Therefore, this paper proposes an algorithm that uses two methods to identify the critical distance relays to be modeled in stability studies: (i) apparent impedance monitoring, and (ii) the minimum voltage evaluation (MVE). The algorithm is implemented in Python 3.6 and uses the GE positive sequence load flow analysis (PSLF) software for performing stability studies. The performance of the algorithm is evaluated on the Western Electricity Coordinating Council (WECC) system data representing the 2018 summer-peak load. The results of the case studies representing various types of contingencies show that to have an accurate assessment of system behavior, modeling the critical distance relays identified by the algorithm suffices, and there is no need for modeling all the distance relays.

*Index Terms*-- Distance relays, identifying critical protective relays, minimum voltage evaluation, power system protection, power system stability, relay misoperation, transient stability study.

## I. INTRODUCTION

THE North American Electric Reliability Corporation (NERC) has identified protection systems as critical reliability assets in modern power grids [1] that play a crucial role in defining the system behavior during and after disturbances. Power system dynamic behavior is governed by two main aspects following a major disturbance: *protection scheme behavior* and *the dynamic characteristics of the fundamental assets of the system including generators, loads, and control devices* [2].

Analyzing the historical data of prior outages and blackouts reveals that unforeseen relay misoperations were one of the main factors in many of these events [3], [4]. Therefore, it is critical to represent protective relays in power system transient stability studies to obtain a realistic assessment of system behavior after a disturbance. Without modeling and assessing the behavior of protection schemes, power system analysis may manifest unrealistic system behavior [5]. For instance, a probable distance relay misoperation after an initial event, which in turn leads to uncontrolled islanding and cascading outages, might not be captured in offline transient stability simulations without including distance relay models [6]. Thus, proper control actions might not be designed to prevent these relay misoperations.

There are different types of protection relays in the system, such as under voltage load shedding (UVLS) and under frequency load shedding (UFLS) relays, that need to be modeled precisely to perform accurate transient stability studies. These protective relays usually are represented in power system stability studies. However, distance relays, which are among the most common and critical protective relays, are usually not modeled in these studies [5]. References [1] and [5]-[7] show that the lack of representation of distance relays in transient stability studies can result in an erroneous analysis of the post-disturbance behavior of power systems. This paper confirms this conclusion via further simulations using the data from the WECC system.

Although modeling all the distance relays in a system for transient stability studies may seem to result in a better assessment of the system behavior [5], modeling thousands of distance relays in a bulk power system, such as the WECC system, is an intractable task. Two main hurdles for proper representation of protective relays in stability studies are addressed in this paper:

- The current limitation of industrial software, such as GE PSLF, does not allow adding thousands of distance relay models to the dynamic file of a bulk power system.
- Protection engineers change the settings of protective relays for various purposes, which makes it a challenging task to ensure updated settings in dynamic models. There are two main settings that govern the functionality in distance relays: (i) the reach of the operation zones, and (ii) the time delay for the operation of each zone. Maintaining updated settings information for thousands of distance relays in dynamic models is an intractable task. Hence, the

This work was supported by Salt River Project (SRP).

Ramin Vakili, Mojdeh Khorsand, and Vijay Vittal are with the School of Electrical, Computer, and Energy Engineering, Arizona State University, Tempe, AZ 85287-5706 USA (e-mail: rvakili@asu.edu; mojdeh.khorsand@asu.edu; vijay.vittal@asu.edu).

Bill Robertson and Philip Augustin are with SRP utilities company, Phoenix, AZ, USA (e-mail Bill.Robertson@srpnet.com; Philip.Augustin@srpnet.com).

setting information of some of the distance relays in the dynamic file of the system, which is used for transient stability studies, might be outdated and different from the actual relay settings. This paper confirms that the outdated relay settings might lead to an invalid assessment of the system behavior during disturbances.

This paper addresses both the challenges by identifying the critical distance relays. The number of these critical relays, that need to be modeled in transient stability studies, is far less than the total number of distance relays in the system.

The need for an algorithm to identify the critical distance relays to be modeled for conducting a precise transient stability study has been identified by the industry [4]-[5] and [8]. In this regard, [9] has developed a link between the computer-aided protection engineering (CAPE) software, which is a software for analyzing protective relay behavior, and the power system simulator for engineering (PSS/E) software. The co-simulation link enables simultaneous analysis of protection systems response and the dynamic response of the system in transient stability studies using the CAPE-Transient Stability (CAPE-TS) module [9]. However, due to a significant computational burden, the developed link in [9] often only analyzes the operation of protection relays in the vicinity of the fault location. In major disturbances, an initiating event may result in unstable power swings, which can cause several distance relay misoperations in the system at locations far from the initial fault location [8]. Therefore, it is essential that any algorithm developed for identifying the critical distance relays be capable of identifying these distance relay misoperations occurring at distant locations as well. Such an algorithm enables a more realistic assessment of power system behavior while using transient stability analysis software, including the CAPE-TS co-simulation platform. Previous research efforts also have developed different methods for identifying the critical protection relays for a contingency. References [6], [10], and [11] have provided approaches for identifying the critical protection relays based on the location and size of the initiating events. However, the approaches used in these references cannot capture the relay operations occurring at the locations far from the fault location. References [12]-[20] have proposed different methods of relay scheme design to distinguish between a power swing and a fault condition. These methods, if properly implemented in conjunction with stability studies, might be able to detect relay operations due to power swings and out-of-step (OOS) conditions. However, they cannot identify all the critical distance relays for performing accurate stability studies. The Independent System Operator (ISO) of New England uses a generic distance relay model that has a zone-3 reach of 300% of line impedances for all the transmission lines. In the planning studies, if the relay trajectories are observed within their zone-3 reach, the actual data of these relays are collected, and the relays are represented in the analysis with accurate settings [21]. The operations of distance relays affect the system response to disturbances. After observing the operation of the first distance relay in a study, the system behavior for the remainder of the simulation time might be different from its behavior without modeling distance relays.

Therefore, the set of critical distance relays, i.e., relays with impedance trajectories within their zone-3 reach, might be different from the set identified in the initial study. Therefore, another study should be performed to identify the new set of critical relays. However, this iterative process for identifying the critical distance relays has not been considered in [21]. Reference [5] has proposed a generic strategy for identifying the critical relays for all contingencies by solving an optimization problem, which is based on prior generator grouping information and network structure. However, this method does not identify critical distance relays that are in the vicinity of the disturbance and may operate due to its initial impacts. Furthermore, the type and location of the disturbance significantly affect the set of critical distance relays to be modeled for conducting accurate transient stability studies. Thus, a methodology that identifies contingency-specific critical distance relays is of interest. The algorithm proposed in this paper utilizes two methods of apparent impedance monitoring and the MVE to identify the critical distance relays for each contingency to enable capturing the local and wide-spread impacts of disturbances.

The rest of this paper is organized as follows. Section II presents the proposed algorithmic approach to identify the critical distance relays. Section III discusses the shortcomings of the existing practices of relay modeling in transient stability studies. Section IV validates the proposed algorithm by analyzing different types of contingencies using the WECC system data. Conclusions are provided in Section V.

## II. IDENTIFICATION OF CRITICAL DISTANCE RELAYS

In this section, first, the two methods used in the proposed algorithm for identifying the critical distance relays are described. Then, the algorithm is explained in detail.

### A. The apparent impedance monitoring method

The apparent impedance monitoring method checks whether the impedance trajectory of any transmission line (seen from both sides of the line) at any time interval of the transient stability study has traversed into any one of the operation zones of its distance relays or not. If that is the case, the distance relays are flagged as critical distance relays.

In order to shed light on the performance of the method, a contingency is studied and the impedance trajectories of some of the distance relays identified by the method are illustrated in Fig. 1. The trajectories from relays at the two ends of a transmission line are shown side-by-side. Fig. 1 shows the characteristics of these critical relays, as well. As seen in this figure, the impedance trajectories of these distance relays traverse into their operation zones. Therefore, the algorithm correctly identifies the distance relays as critical.

### B. Minimum voltage evaluation method

This paper utilizes the analytical MVE method proposed in [22] for detecting unstable power swings, which may lead to distance relay misoperations. The method is based on the fact that during an OOS condition, different groups of generators start to form in the grid. The generators in each group may lose synchronism with respect to the generators in other groups. In this condition, the voltage angle differences at the two ends of some of the lines connecting these groups of generators, i.e. the lines located in the electrical center of the power swing, might become close to 180 degrees, which causes a significant voltage drop along the connecting lines. The distance relays on these lines might misinterpret the situation as a fault and open the lines [22]. Therefore, these distance relays should be considered in transient stability studies since their operations might lead to unintentional islanding and probable cascading outages in the system. The MVE method identifies the lines which are prone to experience distance relay misoperation due to OOS condition. OOS blocking schemes should be included on these lines to block misoperations of their distance relays and avoid uncontrolled islanding. Furthermore, OOS schemes accompany tripping at appropriate locations to initiate controlled islanding during OOS conditions. However, varying operating conditions and disturbances can result in various relay misoperations, which may or may not be equipped with OOS schemes.

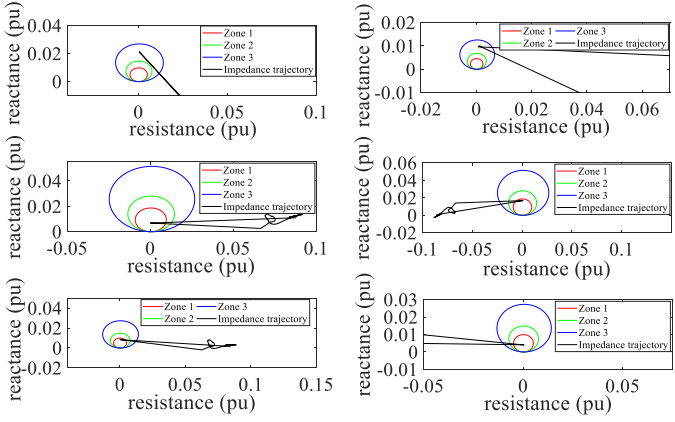

Fig. 1. The impedance trajectories of three of the lines identified by the apparent impedance monitoring method and the operation zone of their relays.

In order to find the lines that may experience voltage drop caused by OOS conditions, the minimum voltage magnitude along all transmission lines are evaluated. The MVE method uses the bus voltages and solves a one-dimensional optimization problem for all the transmission lines at each time step of a transient stability study. If the shunt admittances of the lines are neglected, and the impedances per unit length of the lines are considered to be equal along the line, the minimum voltage along each transmission line at each time step can be obtained by solving the optimization problem (1) and (2).

Objective: Minimize
$|V_a| = \sqrt{((1-a)V_{1r} + aV_{2r})^2 + ((1-a)V_{1i} + aV_{2i})^2}$ (1)
Subject to: $0 \leq a \leq 1$ (2)

In these equations, $V_{1r}$ and $V_{2r}$ are the real parts of the voltages at the two ends of the line. $V_{1i}$ and $V_{2i}$ are the imaginary parts of the voltages at the two ends of the line. $a$ is the fraction of the line length under study and $|V_a|$ is the voltage magnitude at $a$ fraction of the line. The MVE method solves problems (1)-(2) at each step of transient stability study and determines the minimum voltage magnitude along each transmission line. If the minimum voltage along any transmission line at any time step drops to zero without any fault on the line, it can be concluded that the line is along the electrical center and it may experience distance relay misoperation. Thus, distance relays should be modeled on the line for performing transient stability studies. More details of the MVE method are provided in [22].

The contingency of Section 2. A is considered here and the impedance trajectories of some of the lines identified by the MVE method along with their distance relay characteristics are illustrated in Fig. 2. As seen in Fig. 2, the impedance trajectories of these lines (seen from both ends of the lines) traverse into one of the operation zones of their distance relays. Thus, depending on the location of the voltage drop along the line, the relay on either end of the line may respond to the condition by opening the line, which shows that the MVE method has correctly identified the lines along the electrical center.

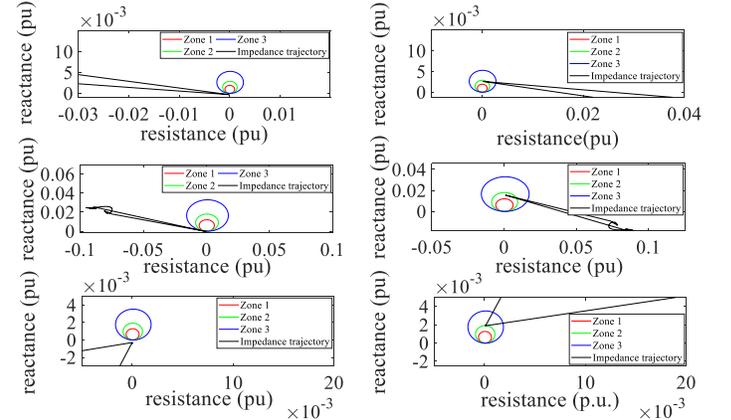

Fig. 2. The impedance trajectories of some of the lines identified by the MVE method and the operation zones of their relays.

*C. The proposed algorithm*

Fig. 3 depicts the flowchart of the proposed algorithm. As shown in this flowchart, in the first step of the algorithm, a transient stability study without modeling any distance relays is performed for the contingency under study. Then, the algorithm obtains all the information required by the apparent impedance monitoring and the MVE methods to identify critical distance relays from the results of the transient stability study. It is to be noted that all the required data including bus voltages and the apparent impedances of the lines are captured at each time interval of the transient stability study without any change in the existing practice. After obtaining the required data from the stability software, the apparent impedance monitoring and the MVE methods identify the transmission lines that are prone to experience distance relay misoperation. Subsequently, in each iteration (except the first), the algorithm checks whether any new critical distance relays, which have not been identified in the last iteration, are identified by either of the two methods. If no new critical distance relay is identified, the algorithm stops; the results of the last iteration of the transient stability study captures the behavior of the system during the contingency. Otherwise, if there exists any new critical distance relay, the algorithm includes their models on both sides of the corresponding transmission lines. Then, the algorithm performs the next iteration of the transient stability study while including the distance relays identified by either of the methods. Then, the algorithm utilizes the two methods again and finds new critical distance relays. This iterative process continues until no new critical distance relay is identified by either of the two methods. Therefore, the proposed algorithm addresses the problems of modeling all the distance relays by significantly reducing the number of distance relays required to be modeled for performing precise transient stability studies. This algorithm equips electric utilities with an efficient offline tool for performing precise transient stability studies for different types of contingencies in their planning studies.

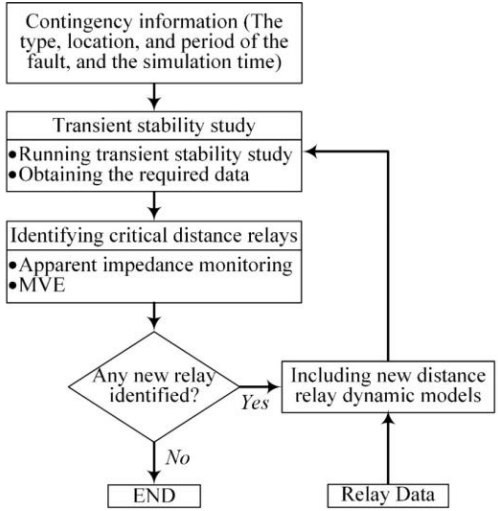

Fig. 3. The flowchart of the proposed algorithm.

## III. DRAWBACKS OF EXISTING RELAY MODELING PRACTICE

Although node-breaker models are gaining more attention in recent years, the common utility practice in transient stability studies is still to represent the system using the bus-branch model. Even though modeling and studying the system using the bus-branch model is easier, it has a drawback associated with simulating bus faults when distance relays are represented. In the bus-branch model shown in Fig.4 (a), when a fault occurs on a bus, all the distance relay models placed on the lines connected to the bus observe the fault and react to it by opening the respective lines. However, this is not usually the case in actual systems, where the substation architecture is in a node-breaker structure as shown in Fig. 4 (b). A bus fault in the node-breaker model is cleared by isolating the faulted bus via breaker opening. Therefore, to obtain a precise characterization of a bus fault, the faulted bus needs to be modeled using a node-breaker model. The closest approximation of an actual bus fault in the bus-branch model is placing a line fault on a small fraction (e.g. 0.05) of the line length from the intended bus. This will approximately have the same effects of a fault on the bus and only the distance relay of one line will react to it. To demonstrate this issue, since we do not have access to the node-breaker representation of the WECC system, a case of bus fault on bus 50 of the small test system shown in Fig. 5 (a) is studied. For comparison, the same bus fault is studied in the bus-branch representation of the test case, shown in Fig. 5 (b). Moreover, a 3-phase line fault on the line between bus 50 and 100 in Fig. 5 (b) is studied in the bus-branch model. The fault is applied at a distance of 5% of the length of the line from bus 50 to approximate the response of the node-breaker model to the bus fault. The faults were cleared after 4 cycles. The rotor angles of generators (with respect to generator 1, the slack generator) for each case are illustrated in Fig. 6. Fig. 6 (a) and Fig. 6 (b) show that the relative rotor angle of generator 2 is different in the two cases of simulating the bus fault using the node-breaker model and the bus-branch model. In Fig. 6 (b), Generator 2 was separated due to multiple relay operations after the initial bus fault. Fig. 6 (c) shows that the line fault occurring at the close vicinity of the bus can approximately imitate the response of the node-breaker model to the bus fault.

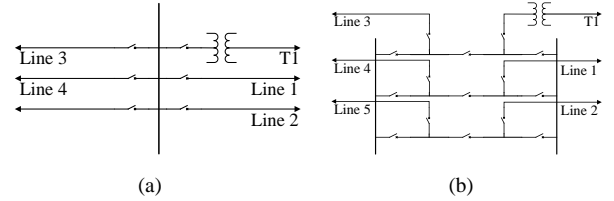

(a)          (b)

Fig.4. the bus-branch (a), and the node-breaker (b) representation of a bus

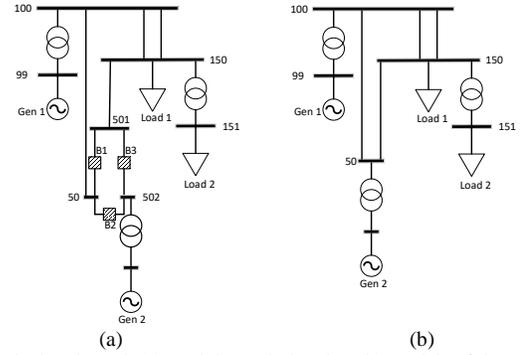

(a)          (b)

Fig.5. the bus-branch (a), and the node-breaker (b) models of the test case

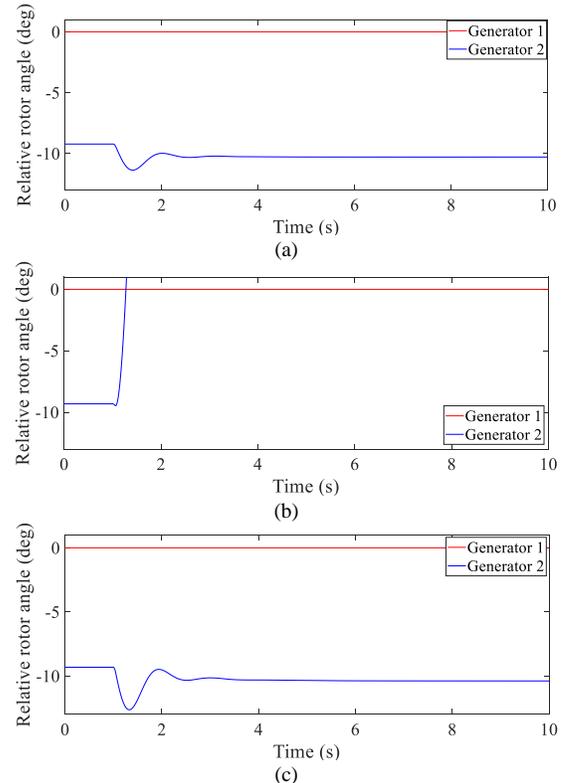

Fig. 6. Relative rotor angles: (a) a bus fault in node-breaker model, (b) a bus fault in bus-branch model, and (c) line fault near a bus in bus-branch model.

Another challenge of relay modeling in stability studies is the nontrivial coordination to maintain and update relay settings for a large number of distance relays. In the current practices of stability studies, for analyzing a disturbance while modeling distance relays, the dynamic models of all of the relays are required to be included. This makes it a challenging task to keep the setting information of distance relays updated in the dynamic models. As mentioned earlier, performing stability studies using outdated relay settings might lead to a completely different assessment of system behavior. To illustrate this point, a line fault is studied on the WECC system. The same line fault is also studied with changes in the zone-2 and -3 reaches of two distance relays from 120 and 220 percent of the respected line impedance to 100 and 120 percent of the line impedance, respectively. A 0.05s change in zone-2 and zone-3 time delays of the distance relays is also applied. The rotor angles of a set of generators are shown in Fig. 7 to show the difference of the system response with the change in the distance relay settings. Fig. 7 confirms that conducting transient stability studies without updated relay settings might lead to invalid results. Therefore, it is necessary to keep track of the changes that are made in distance relay settings. As mentioned in Section 1, the proposed algorithm simplifies this task by significantly reducing the number of distance relays that need to be included in transient stability studies while modeling a disturbance.

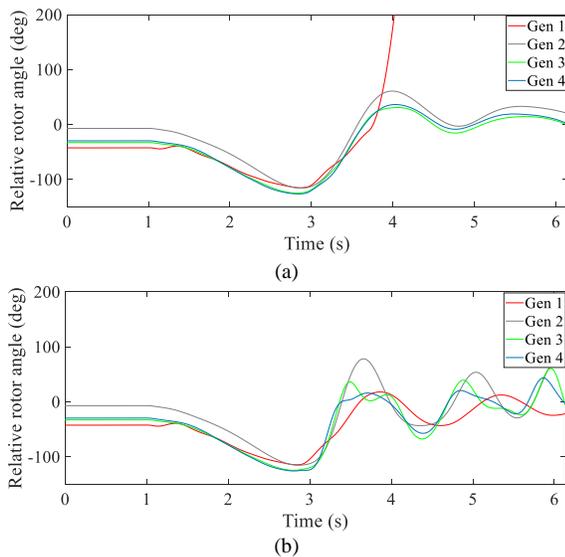

Fig. 7. The relative rotor angles of a set of generators in two cases of (a) using the outdated and (b) using the updated distance relay settings.

## IV. CASE STUDIES

The WECC system data representing the 2018 summer peak load case is used to study the effectiveness of the proposed method. The system includes 23297 buses, 18347 lines, 4224 generators, and 9050 transformers. The maximum generation capacity of the system is 281.38 GW, and the load is 174.3 GW. Throughout this paper, random numbers are used to represent elements, e.g., bus numbers, to protect the proprietary data. The proposed algorithmic approach is tested by comparing its performance in identifying critical distance relays with a reference case, where all the distance relays are included. Two distance relay models from the PSLF model library [23], namely, "Zlin1" and "Zlinw" models, are used in this study:

- "Zlinw", which is a generic distance relay model with two operation zones in the PSLF model library [23], is used to monitor all the lines in the system with the voltage level between 100 kV and 345 kV. This model is included in both the test case and the reference case.
- "Zlin1", which is a more accurate distance relay model with three operation zones in the PSLF model library [23], is used to model distance relays on the transmission lines with the voltage level of 345 kV and above in the reference case. The proposed algorithm is used to identify critical distance relays on transmission lines with the voltage levels of 345 kV and above; Zlin1 models are included on critical distance relay locations during the proposed iterative process. The results of this case with Zlin1 modeled only on critical lines are compared with the reference case, where Zlin1 is modeled on all the lines with the voltage level of 345 kV and above.

It is to be noted that the choice of Zlinw for the lines with the voltage level less than 345 kV is made based on the software limitation as modeling Zlin1 on all the transmission lines overwhelms the dynamic file and is beyond the existing capability of PSLF. The Zlinw model does not need to be modeled for every line with their specific settings; rather, it is just one model with two operation zones that monitors all the lines between its minimum and maximum voltage settings. Although the proposed algorithm is able to identify all the critical lines at any voltage level, in order to compare the results of the algorithm with the reference case, the settings of the algorithm are modified to only identify the critical lines with voltage levels of 345 kV and above. To illustrate the importance of modeling distance relays on the dynamic response of the system, another case is studied that does not include any distance relay models. For all the transmission lines at voltage levels of 345 kV and above in the reference case as well as on the critical lines identified by the algorithm, zone-1, -2, and -3 reaches of distance relay models (Zlin1) are considered to be 80%, 120%, and 220% of the line impedances, respectively. The time delays of zone-1, -2, and -3 operations of distance relay models are set to be 0, 0.2, and 0.3 seconds, respectively. Similar settings are used for zone 1 and 2 of the Zlinw model. Also, the time delay of circuit breakers, i.e. the delay between the time when a distance relay sends tripping signal to a circuit breaker and the time when the circuit breaker opens the line, is set to 0.05 s in both the cases. Note that while generic relay settings are used here for simplicity, the algorithm can be used in planning studies with the accurate relay settings from the protection engineering groups of utilities.

Different types of contingencies including line faults and bus faults followed by removing one or more lines are studied, and the performance of the algorithm is compared with the reference case and the case of not modeling any distance relays.

*Case study 1*: In this case, a line fault is modeled on the 500 kV line between bus 3 to bus 4, which is a part of the California-Oregon Intertie (COI), and is cleared 4 cycles later by opening the line. The results of transient stability studies carried out on the three cases of modeling no distance relays, using the reference dynamic file, and using the proposed algorithm are illustrated in Fig. 8 (a), (b), and (c), respectively. As shown in Fig. 8 (a), without modeling distance relays, only small oscillations are observed in the rotor angles of a set of generators, which are finally damped. However, the same generators start to lose synchronism with respect to each other in Fig. 8 (b), the reference case. Fig. 8 (c) shows that the proposed algorithm is able to capture the behavior of the system exactly like the reference case, even for severe disturbances. While only the relative rotor angles of the selected generators are shown for all case studies for the sake of clarity, the relative rotor angles of all generators show the exact same behavior in the reference case and the proposed method.

Table. I. The list of all distance relay operations in case study 1.

| The reference case | | The proposed algorithm | |
|---|---|---|---|
| Time (s) | Relay | Time (s) | Relay |
| 1.050 | Lines 1, 2, 3 | 1.050 | Lines 1, 2, 3 |
| 1.921 | Line 4 | 1.921 | Line 4 |
| 2.692 | Line 5 | 2.692 | Line 5 |
| 2.862 | Line 6 | 2.862 | Line 6 |
| 2.879 | Lines 7, 8 | 2.879 | Lines 7, 8 |
| 2.900 | Line 9 | 2.900 | Line 9 |
| 3.054 | Line 10 | 3.054 | Line 10 |
| 3.079 | Line 11 | 3.079 | Line 11 |
| 3.108 | Line 12 | 3.108 | Line 12 |
| 3.296 | Line 13 | 3.296 | Line 13 |
| 5.954 | Line 14 | 5.954 | Line 14 |
| 6.291 | Line 15 | 6.291 | Line 15 |
| 6.300 | Line 16 | 6.300 | Line 16 |

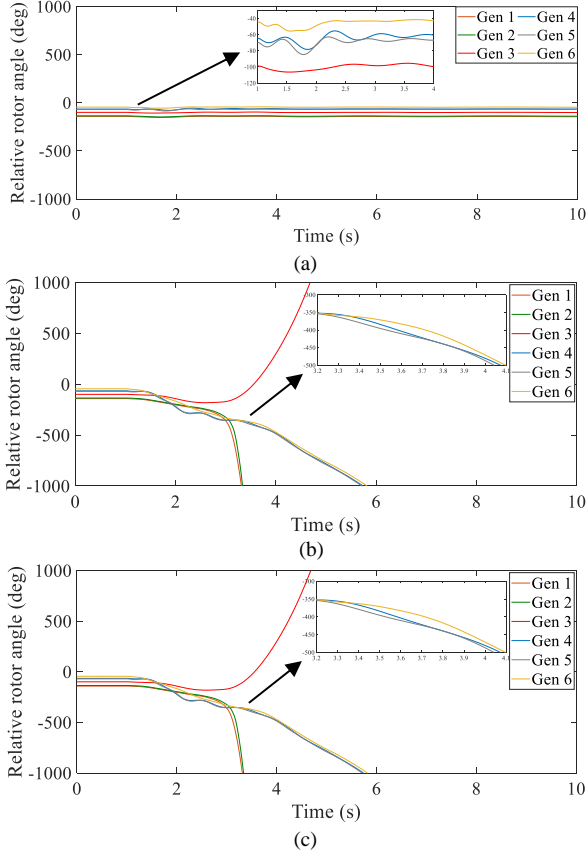

Fig. 8. The relative rotor angles of a set of generators in case study 1: (a) without distance relays, (b) the reference case, and (c) the proposed algorithm.

The disturbance considered has widespread impacts on the system and causes multiple distance relay operations. In a real-world system, there are remedial actions such as controlled islanding and load shedding that start after the occurrence of severe disturbances to alleviate their impacts on the system. However, in this paper, we do not have access to the remedial actions used in the real-word systems, and the transient stability studies are performed without considering them. Nevertheless, real-world remedial actions can be included, and the proposed algorithm is also able to capture the behavior of the system in this case. Table 1 shows the list of the relays which operate in the case of using the reference dynamic file and the proposed algorithm. As seen in Table 1, the same distance relays operate at the same time in both cases, which shows that the algorithm can identify all the critical distance relays in the system.

*Case study 2*: In this case study, a bus fault occurs on bus 3 for 4 cycles and is cleared by removing three COI lines. COI includes three 500 kV transmission lines transferring a total power of around 4113 MW from the north to the south during the summer peak load. These three lines are very critical tie lines of the WECC system since they transfer a huge amount of power during the summer peak load. The outage of these three lines has always been recognized as a critical contingency for the WECC system. Therefore, in this case study, the occurrence of this *N-3* outage after a bus fault on bus 3 is studied. The results of transient stability studies performed in the cases of not modeling distance relays, using the reference dynamic file, and using the proposed algorithm are depicted in Fig. 9 (a), (b), and (c), respectively. Comparing Fig. 9 (a) and Fig. 9 (b) shows that without modeling distance relays, the response of the system is completely different from that of the reference case. Also, Fig. 9 (b)-(c) show that the generator rotor angles are similar in both the reference case and the proposed algorithm, which demonstrates the accuracy of the proposed algorithm in capturing the actual response of the system. Note that with distance relays modeled the network solution in the stability software diverges at 6.283 (s) and the transient stability study terminates after the system becomes unstable in this case study.

For simulating the bus fault, the bus-branch model is used in this paper. However, if the node-breaker model of the system is provided, the algorithm is also expected to be able to identify the critical distance relays in the node-breaker model. The list of the relays, which operate in both cases is given in Table 2. The same distance relays operate at the same time in both cases. This result shows that the proposed algorithm can identify all the critical distance relays in this case as well.

*Case study 3:* To evaluate the performance of the proposed algorithm in unlikely cases of multiple faults occurring consecutively in the system, a case is studied where a line fault occurs on the 500 kV line between bus 3 and bus 5 and is cleared by removing the faulted line and the 500 kV line between bus 3 and bus 6. The second fault occurs 4 seconds later, on the 500 kV line connecting bus 4 to bus 6 and is cleared by removing the faulted line. The results of transient stability studies carried out in the cases of modeling no distance relays, using the reference dynamic file, and using the proposed algorithm are shown in Fig. 10 (a), (b), and (c), respectively. As can be seen in Fig. 10 (a), the dynamic response of the system without modeling distance relays is significantly different from that of the reference case. Comparing Fig. 10 (b) and Fig. 10 (c) shows that the algorithm can accurately capture the behavior of the system during this disturbance. The list of all relay operations in both cases is provided in Table 3. Table 3 shows that the same distance relays operate in both the reference case and the proposed algorithm, which demonstrate that the proposed algorithm accurately identifies all the critical relays. In this case, with modeling distance relays, the network solution diverges at 6.162 (s) and the transient stability study terminates after the system becomes unstable in this case study.

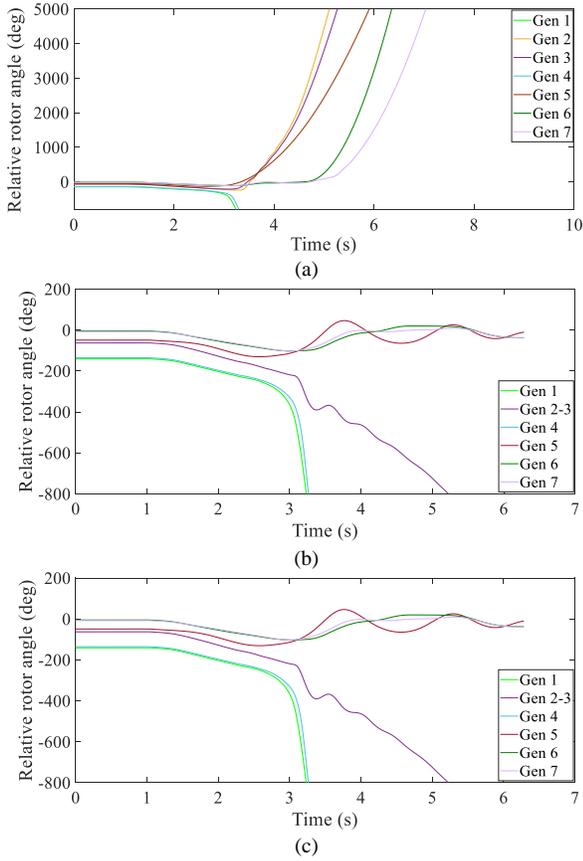

Fig. 9. The relative rotor angles of a set of generators in case study 2: (a) without distance relays, (b) the reference case, and (c) the proposed algorithm.

Table. II. The list of all distance relay operations in case study 2.

| The reference case | | The proposed algorithm | |
|---|---|---|---|
| Time (s) | Relay | Time (s) | Relay |
| 1.050 | Lines 1, 2, 3, 17, 18, 19, 20 | 1.050 | Lines 1, 2, 3, 17, 18, 19, 20 |
| 1.946 | Line 4 | 1.946 | Line 4 |
| 2.750 | Line 5 | 2.750 | Line 5 |
| 2.921 | Line 21 | 2.921 | Line 21 |
| 2.933 | Line 7 | 2.933 | Line 7 |
| 3.025 | Line 8 | 3.025 | Line 8 |
| 3.046 | Line 9 | 3.046 | Line 9 |
| 3.104 | Line 11 | 3.104 | Line 11 |
| 3.117 | Line 10 | 3.117 | Line 10 |
| 3.129 | Line 22 | 3.129 | Line 22 |
| 3.142 | Line 12 | 3.142 | Line 12 |
| 3.321 | Line 13 | 3.321 | Line 13 |
| 5.875 | Line 14 | 5.875 | Line 14 |
| 6.104 | Line 15 | 6.104 | Line 15 |
| 7.712 | Line 23 | 7.712 | Line 23 |
| 8.542 | Line 24 | 8.542 | Line 24 |
| 8.596 | Line 25 | 8.596 | Line 25 |

These case studies show that the proposed algorithm can correctly identify the critical distance relays for different types of contingencies, including the case of occurrence of multiple consecutive faults, and can capture the response of the system after the contingency. However, using the algorithm, distance relays are modeled only on identified critical lines, which noticeably reduces the number of distance relays required to be modeled in comparison with the reference case. Therefore, the proposed algorithm eliminates the problem of modeling all the distance relays in bulk power systems, i.e., software limitations and keeping the relay settings updated in the dynamic files. Table 4 shows the number of distance relays identified by the algorithm as critical in each case study. Table 4 also shows the percentage of the total number of distance relays in the reference case that are identified as critical in each case study. It should be noted that the total number of distance relays in the reference case is 1146. As seen in Table 4, the total number of the critical distance relays (identified by the proposed algorithm) in each case study is less than 10 percent of the total number of distance relays in the reference case. It is worth emphasizing that by identifying only a fraction of the total number of the distance relays that are critical, the proposed algorithm can assist engineers to easily obtain and use the updated relay settings in their analysis to have a more realistic analysis of system behavior. This critical factor greatly reduces model maintenance burden and facilitates better coordination between planning and protection groups of electricity companies.

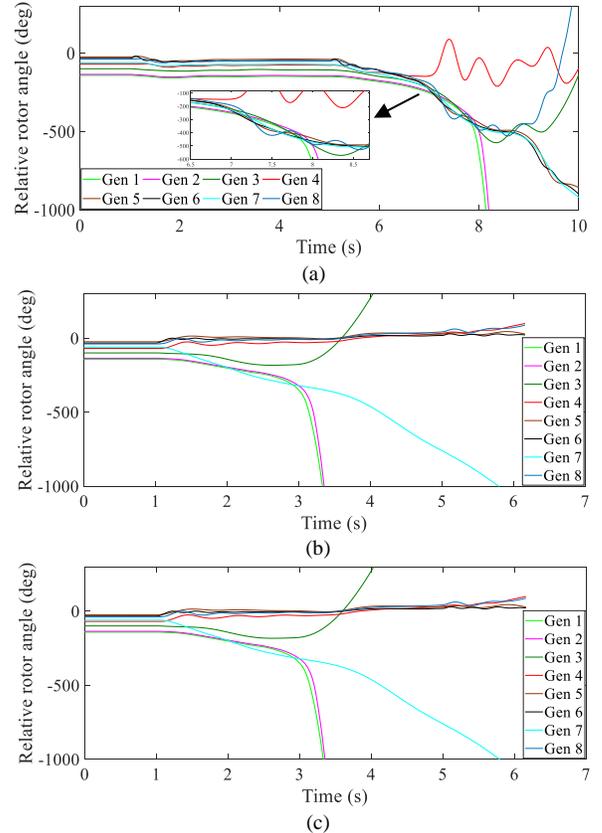

Fig. 10. The relative rotor angles of a set of generators in case study 3: (a) without distance relays, (b) the reference case, and (c) the proposed algorithm.

Table. III. The list of all distance relay operations in case study 3.

| The reference case | | The proposed algorithm | |
|---|---|---|---|
| Time (s) | Relay | Time (s) | Relay |
| 1.050 | Lines 1, 2, 3 | 1.050 | Line 1, 2, 3 |
| 1.933 | Line 4 | 1.933 | Line 4 |
| 2.737 | Line 5 | 2.737 | Line 5 |
| 2.896 | Line 7 | 2.896 | Line 7 |
| 2.904 | Line 21 | 2.904 | Line 21 |
| 2.925 | Line 8 | 2.925 | Line 8 |
| 2.942 | Line 9 | 2.942 | Line 9 |
| 3.010 | Line 10 | 3.010 | Line 10 |
| 3.104 | Line 11 | 3.104 | Line 11 |
| 3.137 | Line 12 | 3.137 | Line 12 |
| 3.321 | Line 13 | 3.321 | Line 13 |
| 5.867 | Line 14 | 5.867 | Line 14 |
| 6.112 | Line 15 | 6.112 | Line 15 |
| 6.162 | Line 16 | 6.162 | Line 16 |

Table IV. The total number of the identified critical distance relays.

| Case study | Total number of the identified critical distance relays | Percentage of the total number of distance relays |
|---|---|---|
| 1 | 84 | 7.33% |
| 2 | 92 | 8.03% |
| 3 | 104 | 9.07% |

## V. CONCLUSION

This paper proposes an algorithm for identifying the critical distance relays that are essential to be modeled for performing accurate transient stability studies of different types of disturbances. The proposed algorithm combines two methods of the apparent impedance monitoring and the MVE to identify the critical distance relays. The algorithm was tested using the WECC system data, and different contingencies including line faults and bus faults were studied to evaluate its performance. The results reveal that transient stability studies carried out without modeling distance relays might not correctly assess the behavior of the system during disturbances. Also, the results demonstrate that the proposed algorithm can precisely identify the critical distance relays required for studying each contingency and can capture the actual behavior of the system. In comparison to the reference case, the proposed algorithm includes a far smaller number of distance relay models in stability studies. Therefore, in bulk power systems like WECC, this method can help in performing stability studies by only modeling the critical distance relays. The results show that conducting transient stability studies using outdated relay settings may lead to an inaccurate assessment of system behavior. However, the algorithm addresses this problem by modeling only critical distance relays for each contingency. Therefore, only a small number of distance relays need to be accurately tracked for changes. Finally, selecting a considerably smaller percentage of distance relays to be modeled in a transient stability study reduces the computational burden of the system and enables the inclusion of more dynamic models in the analysis of various contingencies.